\documentclass[onecollarge,natbib]{svjour2}
\bibpunct{[}{]}{;}{n}{}{,} 
\smartqed  
\usepackage{graphicx}
\usepackage{amsmath,amssymb}
\usepackage{slashed}
\allowdisplaybreaks

\newcommand{\st}{{\scriptscriptstyle T}}
\newcommand{\sL}{{\scriptscriptstyle L}}
\DeclareMathOperator{\tr}{Tr}

\journalname{Few-Body Systems}
\begin{document}

\title{Color effects for transverse momentum dependent parton distribution functions in hadronic processes}
\titlerunning{Color effects for TMDs in hadronic processes}        

\author{M.G.A. Buffing \and P.J. Mulders}


\institute{M.G.A. Buffing \at
				Theory Group, Nikhef \& Department of Physics, Faculty of Science, VU University \\
				De Boelelaan 1081, NL-1081 HV Amsterdam, the Netherlands \\
				\email{m.g.a.buffing@vu.nl}           
				\and
				P.J. Mulders \at
				\email{p.j.g.mulders@vu.nl}}

\date{Received: date / Accepted: date}

\maketitle

\begin{abstract}
In the Drell-Yan process (DY) a quark and an antiquark, coming from two different protons, produce a colorless virtual photon. In the proton description, we include transverse momentum dependent parton distribution functions (TMD PDFs), which go beyond the ordinary collinear PDFs. As such, we become sensitive to polarization modes of the partons and protons that one cannot probe without accounting for transverse momenta of partons, in particular when one looks at azimuthal asymmetries. Matrix elements used in the description of hadronic processes, such as DY, require the inclusion of gauge links, coming from gluon contributions in the process, which are path-ordered exponentials tracing the color flow. In processes with two hadrons in the initial state the color flow between different parts of the process causes entanglement. We show that in the process of color disentangling each gauge link remains sensitive to this color flow. After disentanglement, we find that particular combinations of TMDs require a different numerical color factor than one might have expected. Such color factors will even play a role for azimuthal asymmetries in the simplest hadronic processes such as the DY process.
\end{abstract}

\section{Introduction}
\label{s:intro}
For hadronic interactions, a proper description of the hadrons initiating the process has to be given. One way to do so is by using parton distribution functions (PDFs). We extend this one step further by considering transverse directions as well~\cite{Ralston:1979ys,Tangerman:1994eh,Boer:1999mm}, which give rise to new phenomena, accessible by e.g. manifestations in angular correlations between the particles involved in the process. This way, the PDFs can be extended to transverse momentum dependent PDFs (TMDs). These TMDs are, due to the sensitivity to transverse directions, sensitive to more polarization configurations of the protons and quarks therein compared to PDFs. Furthermore, gluon emissions from the hadrons to the hard processes have to be taken into account. These emissions form gauge links or Wilson lines and are process dependent, as the gluon emission structure will be different for different hard processes. An example where this process dependence plays a role is in the Sivers effect~\cite{Sivers:1989cc,Collins:1992kk,Collins:2002kn,Brodsky:2002rv}. See Ref.~\cite{Bomhof:2006dp} for a tabulation of gauge link structures. A side effect is the introduction of this process dependence into the description of the hadrons and therefore the TMDs, since these gauge links have to be included in their description for a proper color gauge invariant description.

In Sect.~\ref{s:transversemoments}, based on Ref.~\cite{Buffing:2012sz}, we will explain how these gauge links have to be accounted for in the study of quark TMDs. For this, we use transverse moments to make an identification between the TMDs on one side and their corresponding structures in the form of matrix elements on the other side. By identifying the different contributions that are allowed for each TMD, we can use this identification to formulate how the universality of the TMDs is kept (in a more generalized form), while simultaneously allowing for process dependence. For this, it has been shown that the TMDs for a given process are linear combinations of a finite number of universal functions. The formalism in Ref.~\cite{Buffing:2012sz} allows us to catalog all the structures that describe individual correlators. We would like to go one step further, by embedding these correlators in a cross section description of hard interactions, in which case multiple (anti)quark correlators could be present, which have to be described simultaneously. An example of a process where such a situation occurs is Drell-Yan. In describing this process and others, we have to extend the formalism by applying transverse weightings to both correlators simultaneously. In Sect.~\ref{s:colorentanglement}, based on the Refs.\cite{Buffing:2011mj,Buffing:2013dxa}, we write down a first approach to this. The results we present here suggest color entanglement between the correlators.

\section{Including transverse directions}
\label{s:transversemoments}
\begin{figure}[!tb]
\centering
\includegraphics[width=5cm,clip]{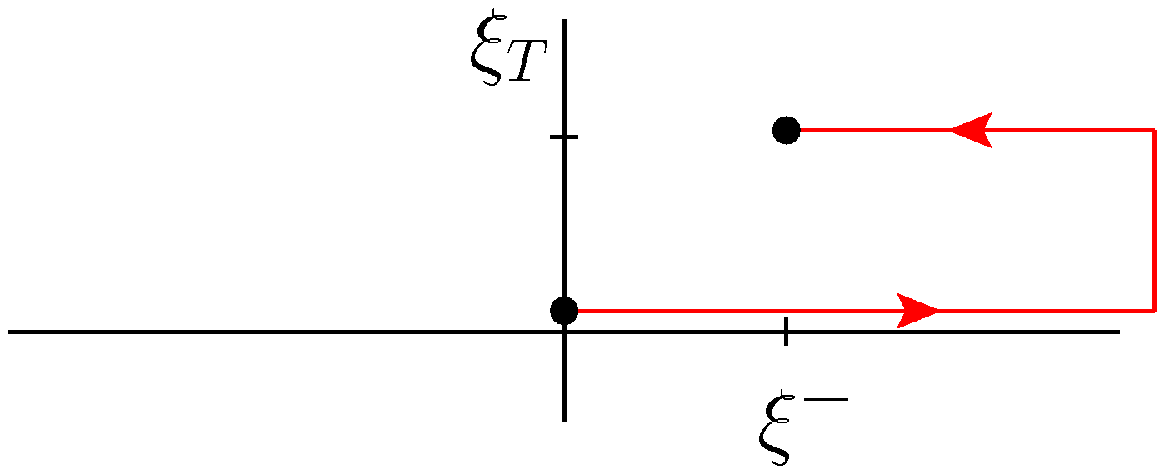}
\hspace{2cm}
\includegraphics[width=5cm,clip]{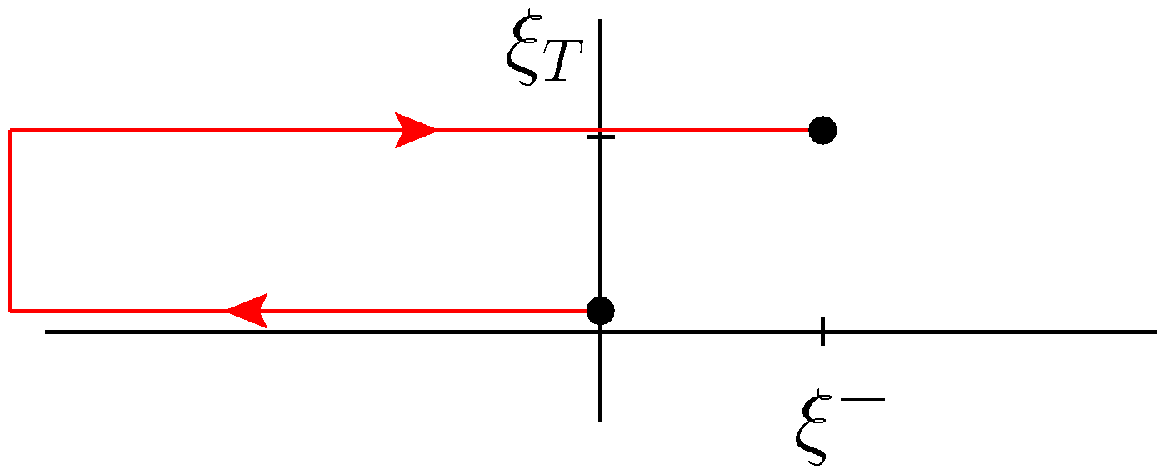}
\\
\begin{small}
(a)\hspace{6.9cm} (b)
\end{small}
\caption{Illustration of the two simplest staple links. In (a) the plus gauge link is illustrated and in (b) the minus gauge link can be seen. The dots indicate the positions of the quark fields and the lines indicate the path between these two fields. Figures taken from Ref.~\cite{Buffing:2012sz}.}
\label{f:GL_quarks}
\end{figure}

Quark correlators can be expanded in a set of transverse momentum dependent parton distribution functions (TMDs) as~\cite{Bacchetta:2006tn}
\begin{eqnarray}
\Phi^{[U]}(x,p_{\st};n)&=&\bigg\{f^{[U]}_{1}(x,p_\st^2)-f_{1T}^{\perp[U]}(x,p_\st^2)\,\frac{\epsilon_{\st}^{\rho\sigma}p_{\st\rho}S_{\st\sigma}}{M}+g^{[U]}_{1s}(x,p_\st)\gamma_{5} \nonumber \\
&& \hspace{2mm}+h^{[U]}_{1T}(x,p_\st^2)\,\gamma_5\,\slashed{S}_{\st}+h_{1s}^{\perp [U]}(x,p_\st)\,\frac{\gamma_5\,\slashed{p}_{\st}}{M}+ih_{1}^{\perp [U]}(x,p_\st^2)\,\frac{\slashed{p}_{\st}}{M}\bigg\}\frac{\slashed{P}}{2}. \label{e:quarkpar}
\end{eqnarray}
In this, we parametrize the spin vector as $S^\mu = S_{\sL}P^\mu + S^\mu_{\st} + M^2\,S_{\sL}n^\mu$ and use the shorthand notation
\begin{eqnarray}
g^{[U]}_{1s}(x,p_T)=S_{\sL} g^{[U]}_{1L}(x,p_{\st}^2)-\frac{p_{\st}\cdot S_{\st}}{M}g^{[U]}_{1T}(x,p_{\st}^2)
\end{eqnarray}
for $g^{[U]}_{1s}$ and $h_{1s}^{\perp [U]}$. Writing the quark correlator using matrix elements we find
\begin{equation}
\Phi_{ij}^{[U]}(x,p_{\st};n) = \int \frac{d\,\xi{\cdot}P\,d^{2}\xi_{\st}}{(2\pi)^{3}}\,e^{ip\cdot \xi}\langle P\vert\overline{\psi}_{j}(0)\,U_{[0,\xi]}\psi_{i}(\xi)\vert P\rangle\,\Big|_{\xi\cdot n=0}. \label{e:quarkmatrix}
\end{equation}
In both the Eqs.~\ref{e:quarkpar} and \ref{e:quarkmatrix}, the $U$'s are gauge links or (a combination of) Wilson lines, path ordered exponentials. For TMDs, these paths are staplelike, illustrated in Fig.~\ref{f:GL_quarks} and given by $U_{[0,\xi]}^{[\pm]} = U_{[0,\pm \infty]}^{[n]}U_{[0_\st,\xi_\st]}^{T}U_{[\pm \infty,0]}^{[n]}$, where $n$ is the light cone direction and $T$ represents the transverse directions. These staplelike gauge links connect the locations of the two quark fields on a path that runs through either minus or plus light cone infinity. More complicated gauge links are allowed as well, see e.g. Refs.~\cite{Bomhof:2006dp,Buffing:2012sz}, but these will always be constructed out of staplelike links.

We would like to carry out an inspection of the matrix element structure of the quark correlator in more detail. For this we use weightings with transverse momenta. At the level of matrix elements, this gives us for the single weighted case~\cite{Boer:2003cm}
\begin{eqnarray}
\Phi_{\partial}^{\alpha[U]}(x) &\equiv & \int d^2 p_{\st}\,p_{\st}^{\alpha}\Phi^{[U]}(x,p_{\st}) = \Big(\Phi_{D}^{\alpha}(x)-\Phi_{A}^{\alpha}(x)\Big)+C_{G}^{[U]}\Phi_{G}^{\alpha}(x) \nonumber \\
&=&\widetilde\Phi_{\partial}^{\alpha}(x) +C_{G}^{[U]}\Phi_{G}^{\alpha}(x), \label{e:Phip}
\end{eqnarray}
where we have to define
\begin{eqnarray}
&& \Phi_D^\alpha(x) = \int dx_1\ \Phi_D^\alpha(x-x_1,x_1\vert x), \label{e:PhiD} \\
&& \Phi_{A}^{\alpha}(x)\equiv \int dx_{1}\,\text{PV}\frac{i}{x_{1}}\,\Phi_{F}^{n\alpha}(x-x_{1},x_{1}|x), \label{e:PhiA} \\[0.2cm]
&& \Phi_G^\alpha(x) = \pi\, \Phi^{n\alpha}_{F}(x,0|x). \label{e:PhiG} 
\end{eqnarray}
In the Eqs.~\ref{e:PhiD}-\ref{e:PhiG}, we have used the definitions
\begin{eqnarray}
\Phi^\alpha_{D\,ij}(x-x_1,x_1\vert x) & = & \int \frac{d\xi{\cdot}P\,d\eta{\cdot}P}{(2\pi)^2}\,e^{ip_1{\cdot}\eta+i(p-p_1)\cdot \xi}\langle P\vert\overline{\psi}_{j}(0) U_{[0,\eta]}\,iD_\st^\alpha(\eta)\,U_{[\eta,\xi]}\psi_{i}(\xi)\vert P\rangle\,\Big|_{LC}, \label{e:PhiD2} \\
\Phi^{\alpha}_{F\,ij}(x-x_1,x_1\vert x) & = & \int \frac{d\xi{\cdot}P\,d\eta{\cdot}P}{(2\pi)^2}\,e^{ip_1{\cdot}\eta+i(p-p_1)\cdot \xi}\langle P\vert\overline{\psi}_{j}(0) U_{[0,\eta]}\,F_\st^{n\alpha}(\eta)\,U_{[\eta,\xi]}\psi_{i}(\xi)\vert P\rangle\,\Big|_{LC}.
\end{eqnarray}
As can be seen, performing transverse weightings of the matrix element in Eq.~\ref{e:quarkmatrix} results in new matrix elements, where additional transverse operators appear in the combinations mentioned above. For the single transverse weighting in Eq.~\ref{e:Phip}, there are two contributions. The $\widetilde\Phi_{\partial}$ will be referred to as partial derivative operator in the rest of this proceeding, while the $\Phi_{G}$ is the gluonic pole contribution (also known as Efremov-Teryaev-Qiu-Sterman matrix element)~\cite{Efremov:1981sh,Efremov:1984ip,Qiu:1991pp,Qiu:1991wg,Qiu:1998ia,Kanazawa:2000hz} that comes with a process dependent gluonic pole prefactor. All gauge link dependence is isolated in this numerical calculable factor. Gluonic poles vanish for fragmentation correlators~\cite{Metz:2002iz,Collins:2004nx,Gamberg:2008yt,Meissner:2008yf,Gamberg:2010uw}, so we focus on distribution correlators only.

Going beyond the single weighted case, we can write down matrix elements with an arbitrary number of gluonic poles and partial derivative contributions. The quark correlator can be described by writing it as the sum of all matrix elements allowed this way as~\cite{Buffing:2012sz}
\begin{eqnarray}
\Phi^{[U]}(x,p_\st) &\ =\ & \Phi(x,p_\st^2)
+ \frac{p_{\st i}}{M}\,\widetilde\Phi_\partial^{i}(x,p_\st^2)
+ \frac{p_{\st ij}}{M^2}\,\widetilde\Phi_{\partial\partial}^{ij}(x,p_\st^2)
+ \frac{p_{\st ijk}}{M^3}\,\widetilde\Phi_{\partial\partial\partial}^{\,ijk}(x,p_\st^2) 
+ \ldots \nonumber \\
&\quad +& C_{G}^{[U]}\bigg(\frac{p_{\st i}}{M}\,\Phi_{G}^{i}(x,p_\st^2)
+ \frac{p_{\st ij}}{M^2}\,\widetilde\Phi_{\{\partial G\}}^{\,ij}(x,p_\st^2)
+ \frac{p_{\st ijk}}{M^3}\,\widetilde\Phi_{\{\partial\partial G\}}^{\,ijk}(x,p_\st^2)
+ \ldots \bigg) \nonumber \\
&\quad +& \sum_c C_{GG,c}^{[U]} \bigg(\frac{p_{\st ij}}{M^2}\,\Phi_{GG,c}^{ij}(x,p_\st^2)
+ \frac{p_{\st ijk}}{M^3}\,\widetilde\Phi_{\{\partial GG\},c}^{\,ijk}(x,p_\st^2)
+ \dots \bigg) \nonumber \\
&\quad +& \sum_c C_{GGG,c}^{[U]}\bigg(\frac{p_{\st ijk}}{M^3}\,\Phi_{GGG,c}^{ijk}(x,p_\st^2)
+\ldots \bigg) + \ldots \, .
\label{e:TMDstructurequarks}
\end{eqnarray}
Some of these terms are summed over the color index $c$, as multiple color configurations exist for a given number of gluonic poles, see e.g. Ref.~\cite{Buffing:2012sz}. For matrix elements containing one gluonic pole at most, only one color configuration appears. Therefore, terms containing a factor of $C_G^{[U]}$ do not have to be summed over color configurations.

As we perform the transverse weightings for the TMDs, the definition
\begin{equation}
f_{\ldots}^{(n)[U]}(x,p_\st^2) = \left(\frac{-p_\st^2}{2M^2}\right)^n\,f_{\ldots}^{[U]}(x,p_\st^2) \label{e:transversemoments}
\end{equation}
is used. Although transverse weightings are strictly speaking defined as the integrated versions of the above functions that therefore do not have any $p_\st$ dependence left, we extend this definition to functions that still have a $p_\st^2$ dependence left, but are integrated over the azimuthal angles.

With the knowledge of transverse weightings at the level of both the matrix elements and TMDs, linking the two descriptions becomes the next hurdle to overcome, for which we use the behavior under time reversal symmetry and the rank of these objects. In the expansion in Eq.~\ref{e:TMDstructurequarks}, matrix elements containing an odd number of gluonic poles are time reversal odd (T-odd), whereas all other terms are time reversal even (T-even). For the TMDs in Eq.~\ref{e:quarkpar} the behavior under time reversal symmetry is known. The TMDs $f_{1T}^{\perp[U]}(x,p_\st^2)$ and $h_{1}^{\perp [U]}(x,p_\st^2)$ are T-odd, while the other functions are T-even. The rank of both matrix elements and TMDs is defined as the number of additional transverse objects they possess, either in the form of prefactors of $p_\st$ or in the form of gluonic poles and partial derivative operators. We give an example for illustrative purposes.

Single transverse weighting of the correlator gives the matrix element structure in Eq.~\ref{e:Phip}, in which both the matrix elements $\widetilde\Phi_{\partial}$ and $\Phi_{G}$ are of rank 1. While $\widetilde\Phi_{\partial}$ is T-even, $\Phi_{G}$ is T-odd. Looking at the TMDs in Eq.~\ref{e:quarkpar}, we see that there are four TMDs which have one prefactor of $p_\st$ which makes them being of rank one. Two of them, $g_{1T}^{[U]}(x,p_\st^2)$ and $h_{1L}^{\perp [U]}(x,p_\st^2)$ are T-even, while the other two, the Sivers function $f_{1T}^{\perp[U]}(x,p_\st^2)$ and the Boer-Mulders function $h_{1}^{\perp [U]}(x,p_\st^2)$ are T-odd. Based on these symmetry arguments we therefore identify $g_{1T}^{[U]}(x,p_\st^2)$ and $h_{1L}^{\perp [U]}(x,p_\st^2)$ with the matrix element $\widetilde\Phi_{\partial}$, while we identify the other two TMDs with the gluonic pole matrix elements. Since the gluonic pole matrix elements come with a gauge link dependent prefactor, we could say that the Boer-Mulders function and the Sivers function come with process dependent prefactors, whereas the two TMDs that are identified with $\widetilde\Phi_{\partial}$ are universal. Generalizing this method, we find that the Pretzelocity function actually consists of three functions, a linear combination of which appearing in any given process, two of which coming with a gluonic pole factor.

\section{Color entanglement}
\label{s:colorentanglement}
In this section, based on the Refs.~\cite{Buffing:2011mj,Buffing:2013dxa}, we go one step beyond the approach in Sect.~\ref{s:transversemoments} by studying the situation where two correlators are considered simultaneously. In this, the (anti)quark correlators might allow for additional complications that do not occur by studying just the individual correlators individually. An example of such a complication is described in Ref.~\cite{Rogers:2010dm}, where it was shown for a specific hadroproduction process that color entanglement occurs. The color entanglement in the specific hadroproduction diagram studied, manifested itself through diagrams with two additional gluons emitted from the correlators. These color charges carrying gluons transferred color between the separate parts of the diagram, as a result of which the diagram could no longer be disentangled and separated into a product of a hard scattering contribution and two quark correlators. In our study we will consider a simpler process. The simplest process with two hadrons in the initial state is the Drell-Yan process, which will be used as illustration.

\begin{figure}[!tb]
\centering
\hspace{0.786195cm}
\includegraphics[width=4.5cm,clip]{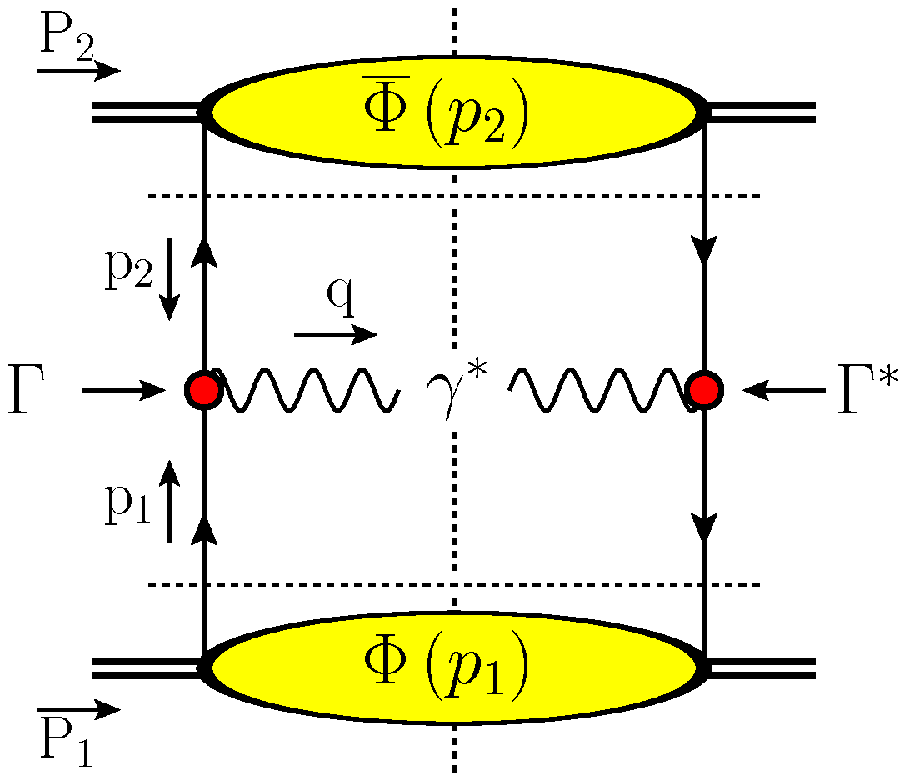}
\hspace{0.786195cm}
\hspace{1cm}
\includegraphics[width=6.07239cm,clip]{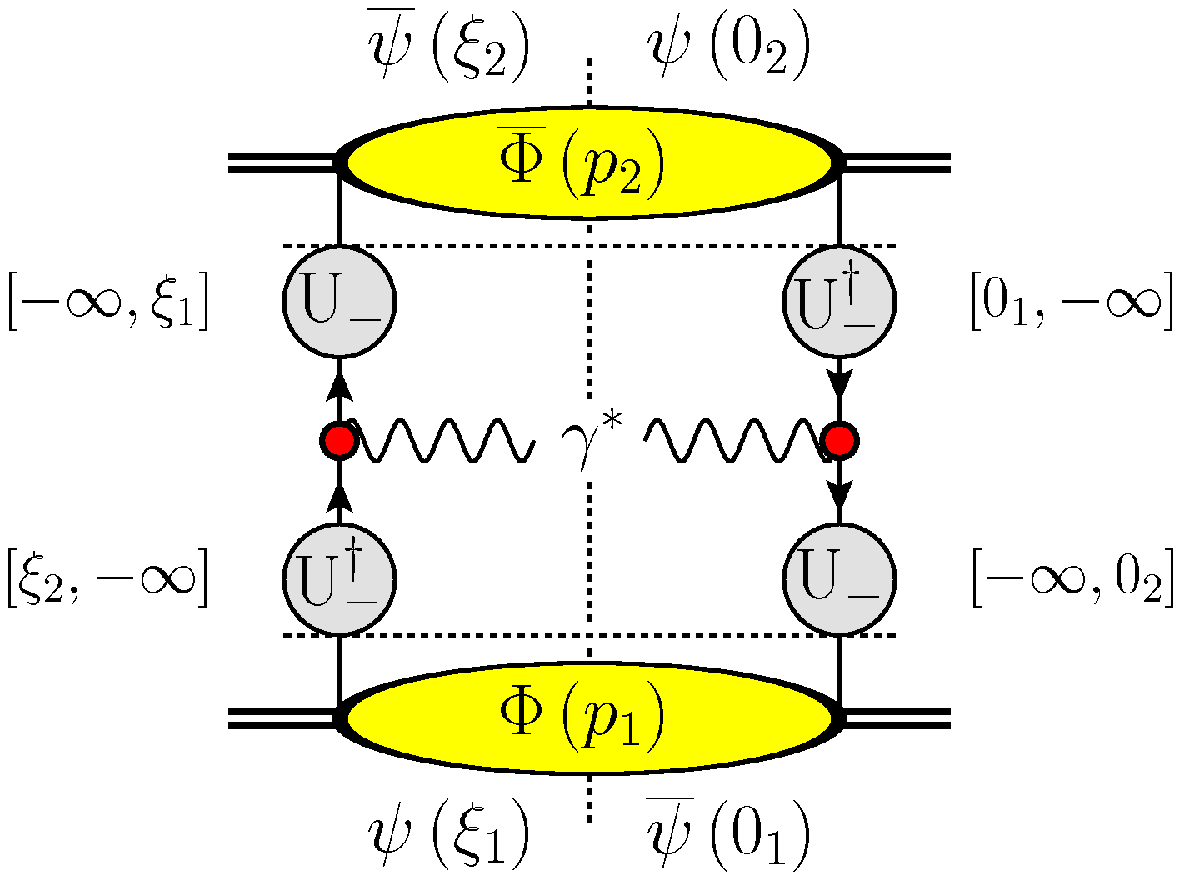}
\\[0.1cm]
\begin{small}
(a)\hspace{6.9cm} (b)
\end{small}
\\[0.1cm]
\caption{The Drell-Yan process illustrated. In (a) the diagram for the cross section can be seen, where the momenta of the relevant particles are indicated. In (b) also the gauge knots are present, where the coordinates next to them indicate the path they take along the light cone direction. Figures taken from Ref.~\cite{Buffing:2013dxa}.}
\label{f:DY}
\end{figure}

In the simplest description, where we omit the gauge links, the Drell-Yan cross section is given by
\begin{eqnarray}
d\sigma_{\text{DY}} & \sim & \tr_c\Big[\Phi(x_1,p_{1\st})\Gamma^{*}\overline\Phi(x_2,p_{2\st})\Gamma\Big] \nonumber \\
& = & \frac{1}{N_c}\Phi(x_1,p_{1\st})\Gamma^{*}\overline\Phi(x_2,p_{2\st})\Gamma, \label{e:basic-0}
\end{eqnarray}
the process itself being illustrated in Fig.~\ref{f:DY}. The gluon emissions, of which the sum of all Feynman diagrams build up the gauge link contributions, couple to the quark and antiquark propagators in the process, illustrated by the blobs in Fig.~\ref{f:DY} with a $U$ in them. The paths in coordinate space of the individual contributions to the gauge links is indicated in square brackets in this figure. We refer to these blobs as gauge knots rather than gauge links, since these gauge knots are not yet evaluated between the quark fields at this point. Writing down the expression for the Drell-Yan cross section including these gauge knots, we find
\begin{eqnarray}
d\sigma_{\text{DY}} & = & \tr_c\Big[U_{-}^{\dagger}[p_2]\Phi(x_1,p_{1\st})U_{-}[p_2]\Gamma^{*} U_{-}^{\dagger}[p_1]\overline\Phi(x_2,p_{2\st})U_{-}[p_1]\Gamma\Big] \label{e:basic-2} \\
& \neq & \frac{1}{N_c}\,\Phi^{[-]}(x_1,p_{1\st})\Gamma^{*}\overline\Phi^{[-^\dagger]}(x_2,p_{2\st})\Gamma . \nonumber
\end{eqnarray}
An entanglement in color space appears at this point, since these gauge knots, nontrivial objects in color space, are not located between the quark fields in the correlators. To study whether this has any effect on the TMD identification process, we again resort to a transverse weighting analysis.

For a direct transverse integration of Eq.~\ref{e:basic-2}, i.e. an integration without including factors of $p_\st$ as weighting no problems occur. The staplelike gauge links reduce to simple light cone gauge links along one direction (the light cone direction). As a result, we find a completely disentangled situation without complications due to the gauge links. The entanglement remains when considering weightings with at least one factor of $p_\st$ for both the partonic momenta involved. We then get
\begin{eqnarray}
\langle p_{1\st}^{\alpha}\,p_{2\st}^{\beta}\sigma_{DY}\rangle =\hspace{3mm}
\frac{1}{N_c}&&\hspace{-3mm}\widetilde\Phi_{\partial}^{\alpha}(x_1)\hspace{0.4mm}\,\Gamma^\ast\,\widetilde{\overline\Phi}{}_{\partial}^{\beta}(x_2)\hspace{0.4mm}\,\Gamma \nonumber \\
- \frac{1}{N_c}&&\hspace{-3mm}\left(\Phi_{G}^{\alpha}(x_1)\,\Gamma^\ast\,\widetilde{\overline\Phi}{}_{\partial}^{\beta}(x_2)\hspace{0.4mm}\,\Gamma +\widetilde\Phi_{\partial}^{\alpha}(x_1)\hspace{0.4mm}\,\Gamma^\ast\,\overline\Phi{}_{G}^{\beta}(x_2)\,\Gamma\right) \nonumber \\
- \frac{1}{N_c^2-1}\frac{1}{N_c}&&\hspace{-3mm}\Phi_{G}^{\alpha}(x_1)\,\Gamma^\ast\,\overline\Phi{}_{G}^{\beta}(x_2)\,\Gamma.
\label{e:DYweight11}
\end{eqnarray}

The color factor in the double gluonic pole term is a result of the color structure in Eq.~\ref{e:basic-2}. The gluonic poles generated through the transverse weightings correspond to a zero momentum gluon emitted by the correlator. In a description in color space there will therefore be color matrices at both sides of this zero momentum gluon. Whenever both the correlators in a process have a gluonic pole contribution, the color matrices of all these contributions become entangled. To describe these terms as a product of two correlators, a color reordering has to be performed. Whenever there is a single gluonic pole for the both the correlators, this reordering gives
\[
\frac{{\rm Tr}_c[T^aT^bT^aT^b]}{{\rm Tr}_c[T^aT^a]\,{\rm Tr}_c[T^bT^b]} = -\frac{1}{N_c^2-1}\frac{1}{N_c}.
\]
This contrast the color factor $1/N_c$ that would have been obtained if the color would not have been entangled. Since a matrix element with a single gluonic pole is identified with a Boer-Mulders function among others, the above situation corresponds to the double Boer-Mulders situation. Therefore, according to this formalism, the double Boer-Mulders function not only has a different color factor multiplying its cross section contribution for Drell-Yan, it appears to contain a sign flip as well.

\section{Conclusions}
\label{s:conclusions}
We use transverse moments to study the universality properties of TMDs. For quarks, this leads to the conclusion that not only the Boer-Mulders function $h_1^{\perp}$ and the Sivers function $f_{1T}$ are sensitive to the process under consideration, the Pretzelocity function $h_{1T}^{\perp}$ is so too. Although the formalism indicates process dependence of the TMDs, it also shows that only a finite number of universal functions exist, with different linear combinations manifesting themselves for different processes. See Ref.~\cite{Buffing:2012sz} for more details.

This mechanism could be extended to cross sections for processes containing multiple (anti)quark correlators, which have to be described simultaneously rather than individually. For Drell-Yan, we find that the transverse momentum weighting formalism as we see it gives us color entanglement at the level of the cross section, which is a result of gauge knots that are trapped in certain parts of the diagram. This color entanglement is of no consequence for most TMD combinations, as it will disappear after a transverse integration. For a few combinations this entanglement leads to additional color factors, an example of which is the double Boer-Mulders combination in the Drell-Yan process, where we get (apart from a different numerical color factor) a sign flip compared to the previously considered result. Further studies will be performed to study this discrepancy between the two results with and without color entanglement.

\begin{acknowledgements}
This contribution is based on a talk given by MGAB. This research is part of the research program of the ``Stichting voor Fundamenteel Onderzoek der Materie (FOM)'', which is financially supported by the ``Nederlandse Organisatie voor Wetenschappelijk Onderzoek (NWO)''. We also acknowledge support of the FP7 EU-programme HadronPhysics3 (contract no 283286) and QWORK (contract 320389). We acknowledge discussions with Asmita Mukherjee, coauthor of Ref.~\cite{Buffing:2012sz}, on which Sect.~\ref{s:transversemoments} is based.
\end{acknowledgements}

\end{document}